\documentclass[reprint,amssymb, amsmath,twocolumn, aps, superscriptaddress, showpacs, footinbib, pra, longbibliography	]{revtex4-2}
\usepackage{amssymb}
\usepackage{amsmath}
\usepackage{color}
\usepackage[colorlinks,bookmarks=false,citecolor=blue,linkcolor=blue,urlcolor=blue]{hyperref}
\usepackage{graphicx}
\usepackage{footmisc,booktabs}
\usepackage{gensymb}
\usepackage{sidecap}
\usepackage{microtype}
\usepackage{epstopdf}

\begin{document}

\title{Dimensional Crossover and Correlated Spin Dynamics in the Chain-Like van der Waals Antiferromagnet AgCrP$_2$S$_6$ }

\author{C. S. Saramgi}
\affiliation{Leibniz Institute for Solid State and Materials Research Dresden, 01069 Dresden, Germany}
\affiliation{Institute for Solid State and Materials Physics, TU Dresden, 01062 Dresden, Germany}
\author{Y. Shemerliuk}
\affiliation{Leibniz Institute for Solid State and Materials Research Dresden, 01069 Dresden, Germany}
\affiliation{Institute for Solid State and Materials Physics, TU Dresden, 01062 Dresden, Germany}
\author{S. Selter}
\affiliation{Leibniz Institute for Solid State and Materials Research Dresden, 01069 Dresden, Germany}
\author{S. Aswartham}
\affiliation{Leibniz Institute for Solid State and Materials Research Dresden, 01069 Dresden, Germany}
\affiliation{International Research Centre MagTop, Institute of Physics, Polish Academy of Sciences, al. Lotnik\'{o}w 32/46, 02-668 Warsaw, Poland}
\author{B. Büchner}
\affiliation{Leibniz Institute for Solid State and Materials Research Dresden, 01069 Dresden, Germany}
\affiliation{Institute for Solid State and Materials Physics, TU Dresden, 01062 Dresden, Germany}
\affiliation{Würzburg-Dresden Cluster of Excellence ct.qmat, Technische Universität Dresden, 01069 Dresden, Germany}
\author{H.-J. Grafe}
\email{H.grafe@ifw-dresden.de}
\affiliation{Leibniz Institute for Solid State and Materials Research Dresden, 01069 Dresden, Germany}
\author{K. M. Ranjith}
\email{ranjith.km1857@gmail.com}
\affiliation{Center for Quantum Technologies and Complex Systems (CQTCS), Christ University, Bengaluru, Karnataka 560029, India}
\affiliation{Leibniz Institute for Solid State and Materials Research Dresden, 01069 Dresden, Germany}
\affiliation{Würzburg-Dresden Cluster of Excellence ct.qmat, Technische Universität Dresden, 01069 Dresden, Germany}

\date{\today}

\begin{abstract}\noindent
We report a comprehensive investigation of the magnetic properties and spin dynamics of the chain-like van der Waals antiferromagnet AgCrP\(_2\)S\(_6\) using magnetization, magnetic susceptibility, and \(^{31}\)P nuclear magnetic resonance (NMR). The magnetic susceptibility exhibits a broad maximum near \(T_{\rm max}\approx 250\) K, characteristic of strong short-range antiferromagnetic correlations in a low-dimensional system. A classical Fisher-chain analysis gives an effective intrachain exchange scale \(J/k_B\approx140\) K, while a quantum \(S=3/2\) nearest-neighbor-chain calculation yields \(J/k_B\approx114\) K. A constrained AF–FM zig-zag-chain calculation is also compatible with the susceptibility, supporting a dominant quasi-one-dimensional antiferromagnetic exchange scale of order \(10^2\) K.

 Low-temperature magnetization measurements reveal pronounced magnetic anisotropy, with a spin-flop transition near \(\mu_0H_{\rm sf}\approx 6.5\) T for fields applied along the \(b\)-axis, identifying the easy axis of the ordered state.

\(^{31}\)P NMR measurements provide microscopic insight into both static and dynamic spin correlations. The paramagnetic spectra exhibit a Pake doublet arising from dipolar coupling within the structural P–P dimer. The \(K\)–\(\chi\) relation displays two distinct linear regimes separated near \(T^\ast\approx 250\) K, with different slopes indicating a change in the effective transferred hyperfine couplings. At high temperatures, the spin-lattice relaxation rate is nearly temperature independent and significantly exceeds the exchange-narrowed estimate based on the scalar hyperfine coupling. This discrepancy indicates that the relaxation is governed by the full anisotropic hyperfine form factor, including transverse and possibly off-diagonal tensor components, together with finite-\(q\) spin fluctuations. Upon cooling toward the Néel temperature \(T_{\rm N}\approx 20\) K, the relaxation rate increases due to the growth of spin correlations and exhibits critical enhancement near the transition. Below \(T_{\rm N}\), the relaxation follows \(1/T_1\propto T^5\), consistent with magnon-mediated relaxation in the ordered state. The temperature dependence of the internal hyperfine field yields a critical exponent \(\beta\approx 0.36\), consistent with three-dimensional universality. 

Together, these results establish AgCrP\(_2\)S\(_6\) as a quasi-one-dimensional \(S=3/2\) antiferromagnet embedded in a layered van der Waals host, in which local probes directly resolve the crossover from correlated intrachain spin dynamics to anisotropic three-dimensional magnetic order. More broadly, AgCrP\(_2\)S\(_6\) provides a useful van der Waals platform for studying dimensional crossover, correlated spin dynamics, and hyperfine-coupling evolution in low-dimensional antiferromagnets.
\end{abstract}

\maketitle
\section{Introduction}

Low-dimensional quantum magnets provide an exciting platform for
exploring correlation-driven phenomena that deviate from conventional
three-dimensional (3D) magnetic behavior~\cite{DagottoRice1996,Giamarchi2004}. In one-dimensional (1D)
systems, thermal and quantum fluctuations strongly suppress long-range
magnetic order, leading instead to extended regimes dominated by
short-range spin correlations and collective excitations~\cite{MerminWagner1966,Giamarchi2004}.
When weak interchain interactions are present, these correlated
spin chains may eventually develop long-range magnetic order at a
temperature $T_\mathrm{N}$ much smaller than the dominant intrachain exchange
interaction $J$, producing a dimensional crossover from 1D correlated
behavior to 3D magnetic order~\cite{Schulz1996,Yasuda2005}. Understanding how such correlated
low-dimensional spin fluctuations evolve into long-range order
remains a central problem in condensed matter physics.

Theoretical studies of quasi-1D Heisenberg
antiferromagnets show that the onset of long-range order is controlled
by weak interchain coupling $J_\perp$.
Within random-phase approximation (RPA) approaches, the ordering condition is
$J_\perp \chi_{1D}(Q,T_\mathrm{N}) \sim 1$,
where $\chi_{1D}$ is the susceptibility of an isolated spin chain at
the ordering wave vector $Q$~\cite{Schulz1996,Yasuda2005}.
Consequently, even very small interchain interactions can stabilize
3D magnetic order once the intrachain correlation
length becomes sufficiently large.
The ratio $T_\mathrm{N}/J$ therefore provides a useful measure of the exchange
anisotropy and dimensionality of the magnetic interactions~\cite{Schulz1996,Yasuda2005}.
Experimental realizations of quasi-1D antiferromagnets
span a wide range of spin values, from strongly quantum
$S=1/2$ systems such as Sr$_2$CuO$_3$ and KCuF$_3$~\cite{Motoyama1996,lake2005,Tennant1995}
to higher-spin compounds where quantum fluctuations are reduced
but pronounced dimensional crossover effects persist~\cite{DeFotis1990,Kenzelmann2002}. At sufficiently low energies, half-integer antiferromagnetic spin chains may also be described within the Tomonaga–Luttinger-liquid framework, which has been investigated experimentally in several quasi-one-dimensional magnetic systems.~\cite{Klanj2015,Faure2019}

Layered van der Waals (vdW) magnets have recently emerged as a
versatile platform for realizing anisotropic magnetic interactions
and reduced dimensionality \cite{Huang2017,Gong2017,Burch2018}.
The weak interlayer bonding in these materials enables strong
directional dependence of exchange pathways within the layers and
provides opportunities to tune magnetic properties through strain,
pressure, or thickness reduction~\cite{Siskins2022,Cenker2021}. The ability to isolate atomic layers and construct heterostructures makes 
vdW magnets particularly attractive for fundamental studies of low-dimensional 
magnetism and potential applications in magnonic devices and quantum 
technologies~\cite{Baltz2018,Chumak2015}.
While much attention has focused on two-dimensional (2D) vdW magnets with
honeycomb lattices~\cite{Abraham2023,Senyk2023,van2024,mayorga2017,cheng2021,Wildes2015,upreti2025,Lee2016}, comparatively fewer systems host intrinsically
chain-like magnetic networks embedded within layered structures.
Such materials combine quasi-1D exchange anisotropy
with the structural flexibility of vdW crystals, making them
particularly attractive for investigating dimensional crossover
phenomena~\cite{Basnet2024}.

\begin{figure*}[!htp]
  \centering
  \includegraphics[clip,width= 2\columnwidth]{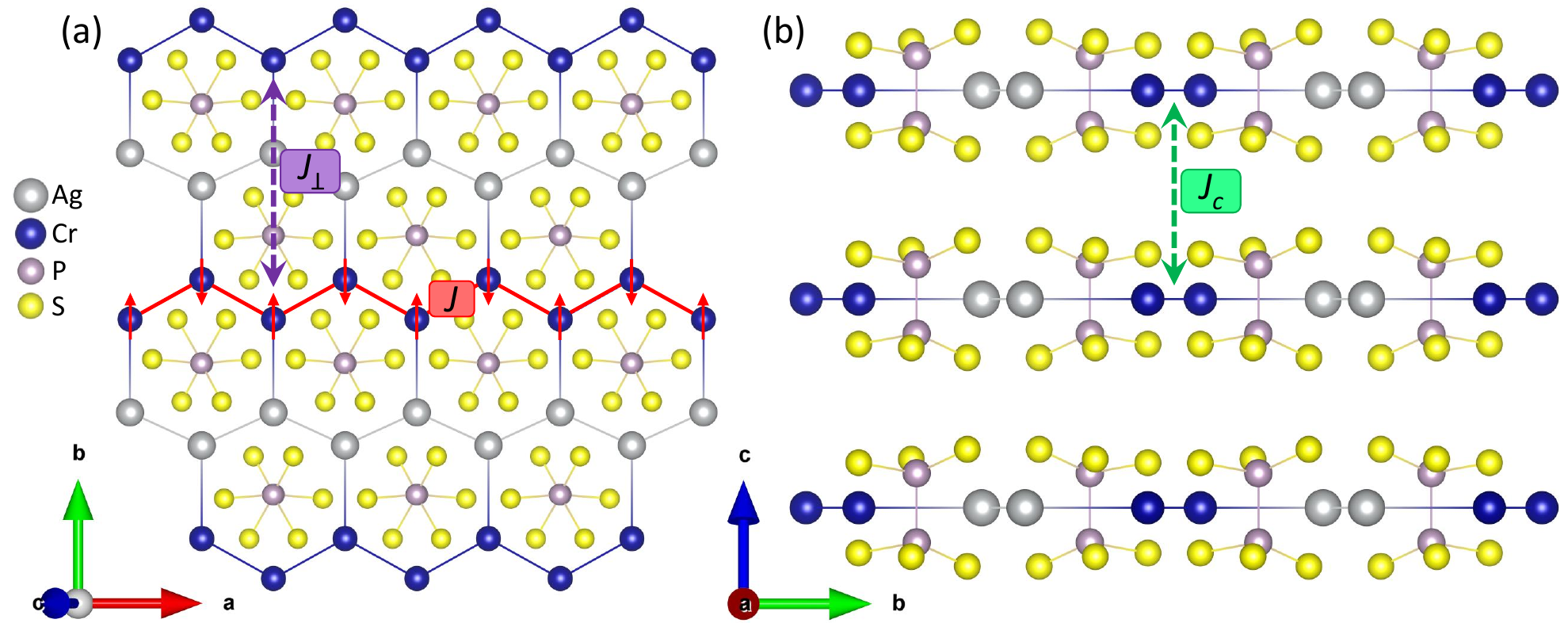}
  \caption{Crystal structure and schematic magnetic exchange topology of AgCrP\(_2\)S\(_6\). (a) View normal to the \(ab\) plane, showing the chain-like arrangement of Cr\(^{3+}\) ions along the crystallographic \(a\) direction. The red arrows over $\mathrm{Cr}^{3+}$ ions symbolise their spin orientation. The dominant intrachain exchange \(J\) and weaker in-plane interchain coupling \(J_\perp\) are indicated schematically. (b) Side view of the stacked van der Waals layers, with \(J_c\) denoting the weaker interlayer coupling.}\label{structure}
\end{figure*}

AgCrP$_2$S$_6$ belongs to the family of  metal
thiophosphates related to the $M$P$X_3$ compounds~\cite{Selter2021,Basnet2024}.
It crystallizes in the monoclinic space group $P2/a$ and adopts a layered vdW structure. Within each layer, the Ag and Cr atoms occupy a honeycomb-like lattice. The magnetic Cr$^{3+}$ ions ($S=3/2$) are coordinated by sulfur in distorted octahedral environments
within each layer and the layers are separated by van der Waals
gaps.  A notable feature of this compound is the pronounced anisotropy of
Cr–Cr distances. Along the crystallographic $a$-axis, the Cr ions are separated by a
short distance of approximately $3.3$~\AA, forming well-defined
magnetic chains, whereas the in-plane separations perpendicular to
the chains are significantly larger ($\sim9.5$~\AA).
The interlayer separation is intermediate ($\sim6.7$~\AA),
reflecting the layered van der Waals structure~\cite{Selter2021}.
This hierarchy of distances naturally favors dominant intrachain
superexchange interactions along the short bonds and much weaker
interchain and interlayer couplings. The resulting hierarchy of magnetic interactions is illustrated schematically in Fig.~\ref{structure}, where the dominant intrachain exchange \(J\) along the crystallographic \(a\) direction is distinguished from the weaker interchain \(J_\perp\)  and interlayer couplings \(J_c\). This quasi-one-dimensional exchange topology, embedded within a layered van der Waals lattice, makes AgCrP\(_2\)S\(_6\) a particularly appealing platform for investigating how low-dimensional spin correlations evolve into long-range order.

The magnetic exchange topology can therefore be approximated as a set of antiferromagnetic chains weakly coupled within a layered vdW framework. Bulk susceptibility measurements reveal a broad maximum and long-range antiferromagnetic order below 
$T_\text N\approx$ 20~K~\cite{Payen1992,Mutka1993,Lee2025,park2024}. In the present work, we extend the temperature range of the susceptibility measurements and resolve the magnetic response along all three principal crystallographic directions. The large separation between the exchange scale $J$ and the ordering temperature suggests substantial suppression of long-range order by low-dimensional correlations, consistent with the structural exchange anisotropy.

The related $S=1$ chain compound AgVP$_2$S$_6$ realizes the integer-spin Haldane limit with a large spin gap~\cite{Mutka1989,Asano1994}, whereas AgCrP$_2$S$_6$ provides the half-integer $S=3/2$ counterpart in which weak interchain coupling drives long-range antiferromagnetic order.

Despite these indications, several issues remain unresolved. First, the temperature scale at which short-range intrachain correlations become significant has not been identified. Second, the extent to which correlated spin dynamics persist above $T_\mathrm{N}$ in this intermediate-spin ($S=3/2$) regime is unclear. Third, the nature of the critical behavior and low-temperature excitations in the ordered state has not been established experimentally. Resolving these issues is essential for establishing whether AgCrP$_2$S$_6$ is simply another anisotropic antiferromagnet or a broader van der Waals platform for dimensional crossover physics.

Nuclear magnetic resonance (NMR) is a powerful microscopic probe
for addressing these questions. The NMR shift directly measures the local magnetic susceptibility
at wave vector $q=0$~\cite{Slichter1990}, while the spin–lattice relaxation rate $(1/T_1)$ probes dynamical spin fluctuations integrated over the Brillouin zone~\cite{Moriya1956,Moriya1963}.
Deviations from linear $K$–$\chi$ scaling can signal the redistribution
of spectral weight away from the uniform susceptibility, but may also reflect an evolution of the effective hyperfine coupling when the nucleus samples a correlated sum of transferred hyperfine fields from multiple neighboring moments.
Furthermore, when nuclear sites form structural dimers coupled to
the same magnetic ions, correlated hyperfine-field fluctuations can
significantly modify the nuclear relaxation dynamics
\cite{Beeman1968,Slichter1990}. 
Thus, combined analysis of static and dynamic NMR observables
provides detailed insight into both spatial and temporal aspects of
spin correlations as demonstrated in related $T_2$P$_2$S$_6$ materials~\cite{Dioguardi2020,Bougamha2022,Saramgi2026}.


In this work, we investigate the magnetic correlations and spin dynamics of AgCrP$_2$S$_6$ using bulk magnetization and
$^{31}$P NMR measurements. The results show that the system exhibits strong intrachain antiferromagnetic
correlations with an exchange interaction of order \(10^2\) K, leading to a pronounced separation between the exchange scale and
the ordering temperature in good agreement with earlier susceptibility and DFT estimates of the intrachain exchange~\cite{Mutka1993,park2024}. 
NMR-shift and relaxation measurements reveal the development of short-range correlations well above $T_\mathrm{N}$, while the ordered phase is characterized by anisotropic 3D critical behavior and magnon-mediated spin dynamics. 
Taken together, these results show that AgCrP$_2$S$_6$ is a van der Waals antiferromagnet in which local probes track the evolution of spin correlations across a pronounced dimensional crossover, involving both a redistribution of spectral weight from $q=0$ to finite-$q$ antiferromagnetic correlations and a change in the effective hyperfine coupling.

\section{EXPERIMENTAL}

Single crystals of AgCrP$_2$S$_6 $ were grown by the Chemical Vapor Transport (CVT) using silver, chromium, red phosphorus, and sulfur as precursors, with iodine serving as the transport agent. Shiny, flat, plate-like structures were obtained, among which a crystal measuring \( 2 \times 1 \times 0.28 \, \text{mm} \) was studied in this work. The technique for crystal growth and the associated characterization methods are discussed elsewhere~\cite{Selter2021}. 
The DC magnetization of the single-crystal sample was measured using the Quantum Design SQUID magnetometer (MPMS3). Data were collected with the magnetic field applied along the three principal crystallographic directions ($a, b$, and $c^*$). Magnetic susceptibility ($\chi(T)$) was measured at a fixed field of 5~T, and isothermal magnetization $M(H)$ was recorded at 2~K.

The NMR measurements on AgCrP$_2$S$_6$ single crystals were conducted in Magnex Scientific and Oxford superconducting magnets across various magnetic fields. The systems employed a home-built probe with a single-axis goniometer for precise sample orientation along the crystallographic axes. All NMR experiments were performed using a Tecmag Apollo spectrometer. $^{31}\text{P}$ nuclei (nuclear spin $I= 1/2$, gyromagnetic ratio $^{31}\gamma/2\pi$ = 17.2356 MHz/T and $100\%$ natural abundance) were utilized to probe the sample's local environment. A standard spin-echo pulse sequence $(\pi/2 - \tau - \pi)$ with $\tau$ = 10~$\mu$s was used for spectral measurements, and the spin-lattice relaxation rates $(T_{1}^{-1})$ were determined using a saturation-recovery pulse sequence $(\pi/2 -\pi/2 - \tau -  \pi)$. NMR shift was calculated with reference to the resonant frequency of the bare nucleus $^{31}$P, $\nu_{\text ref} =(^{31}\gamma/2\pi)\mu_0\textbf{H}$, using
$K(T)=\frac{\nu(T)-\nu_{\mathrm{ref}}}{\nu_{\mathrm{ref}}},$
where $\nu(T)$ is the observed resonance frequency. The magnetic field was calibrated  from the NMR signal from the copper coil. Samples were oriented along the different crystallographic axes via a room-temperature rotation experiment. Temperature-dependent measurements were then conducted down to 7 K using a flow cryostat (with the sample in a helium gas atmosphere) and a calibrated Lakeshore Cernox temperature sensor.

\section{RESULTS}

\subsection{Magnetization and magnetic susceptibility}

\begin{figure}[!htp]
  \centering
  \includegraphics[clip,width=1\columnwidth]{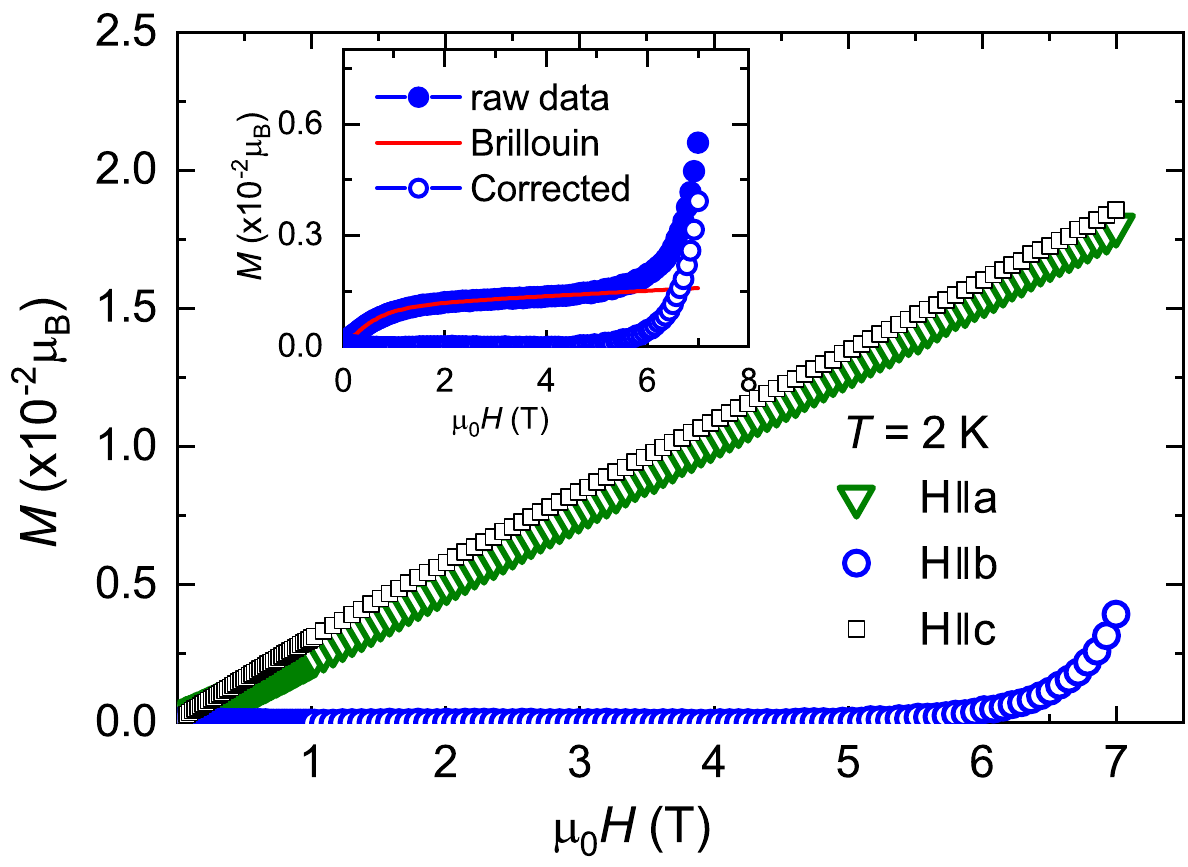}
  \caption{Isothermal magnetization $M(H)$ measured at $T=2$~K for
magnetic fields applied along the three principal crystallographic
directions.  The inset shows the raw data
together with the fitted Brillouin contribution. The main panel shows
the magnetization after subtraction of the Brillouin contribution. }\label{fig:MH}
\end{figure}

Figure~\ref{fig:MH} shows the isothermal magnetization $M(H)$ measured at
$T=2$ K for magnetic fields applied along the three principal crystallographic
directions. For $\mu_0H \parallel a$ and $\mu_0H \parallel c^*$, the magnetization increases nearly
linearly over the entire field range up to 7~T, indicating that the system remains far
from magnetic saturation in the accessible field range. In contrast, $M(H)$
for $\mu_0H \parallel b$ exhibits a pronounced curvature at low fields, followed by a change
in slope near $\mu_0H \approx 6.5$~T. This anomaly corresponds to a spin-flop transition
characteristic of a uniaxial antiferromagnet and identifies the $b$-axis as the magnetic
easy axis.

The low-field curvature observed in the raw $M(H)$ data originates from a small
paramagnetic impurity contribution superimposed on the intrinsic antiferromagnetic
response. To quantify this contribution, the magnetization was modeled as

\begin{equation}
M(H)=\chi H + n_{\mathrm{imp}} g\mu_B S
B_{3/2}\!\left(\frac{g\mu_B S H}{k_B (T+\theta_{\mathrm{imp}} )}\right),
\end{equation}

where $\chi H$ represents the intrinsic linear antiferromagnetic background,
$B_{3/2}(x)$ is the Brillouin function for $S=3/2$, $n_{\mathrm{imp}}$ is the fraction
of dilute free spins, and $\theta_{\mathrm{imp}} $ accounts phenomenologically for weak interactions
between impurity moments. A satisfactory fit to the low-field region yields
$n_{\mathrm{imp}} \approx 3.5\times10^{-4}$ (corresponding to $\sim0.035\%$ free spins per
Cr) and $\theta_{\mathrm{imp}}  \approx -0.9$ K. Subtracting the resulting Brillouin contribution
removes the impurity-induced curvature and reveals the intrinsic magnetization, which
is nearly linear for $H \parallel a$ and $H \parallel c^*$, while for $H \parallel b$ the
corrected curve clearly displays the spin-flop transition near 6.5~T. Because the
impurity spins are dilute and nearly isotropic, the same Brillouin contribution was
subtracted from the magnetization for all field orientations.

\begin{figure}
  \centering
  \includegraphics[clip,width=1\columnwidth]{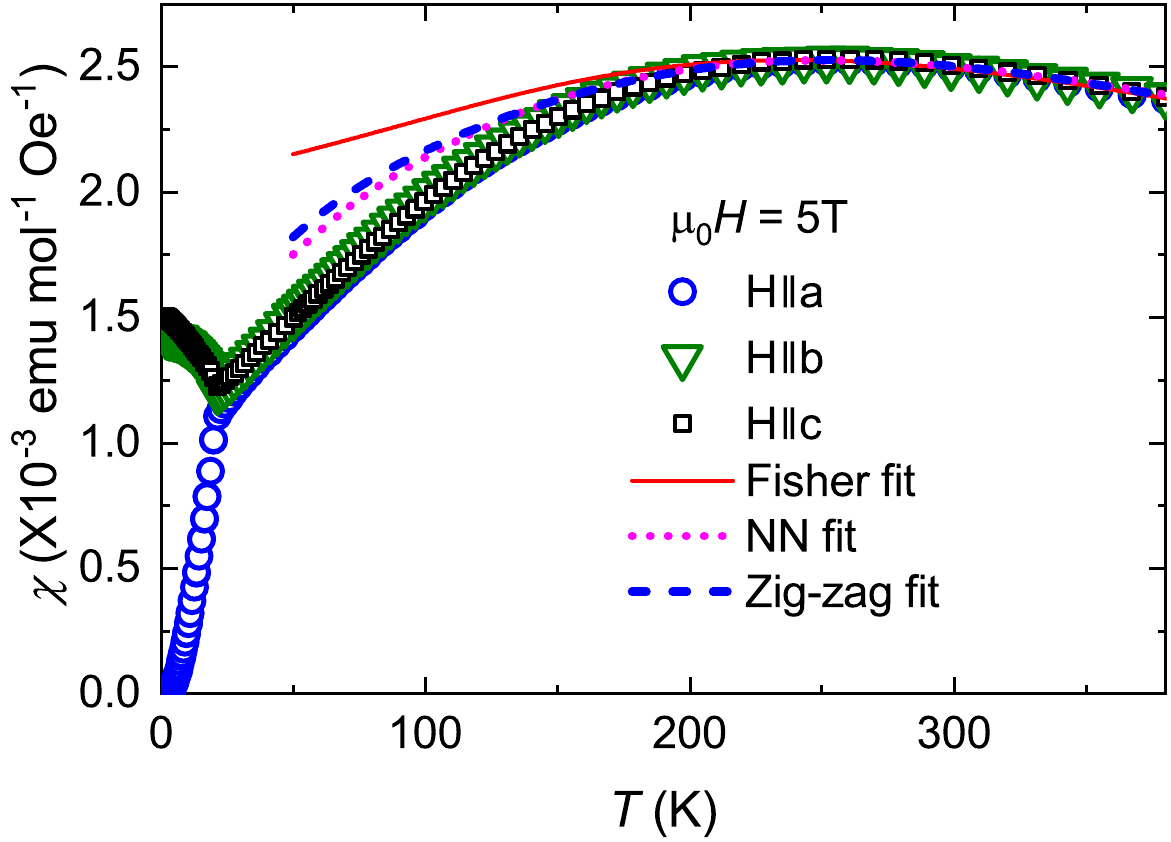}
  \caption{Temperature dependence of the magnetic susceptibility $\chi(T)$ measured
in an applied field of $\mu_0H=5$ T for different field orientations.
The data are shown after subtraction of the paramagnetic impurity
contribution obtained from the modified Brillouin fit to the
magnetization data. The curves show the classical Fisher-chain, quantum \(S=3/2\) nearest-neighbor-chain, and the constrained AF–FM zig-zag-chain descriptions discussed in the text.}\label{fig:chiT}
\end{figure}

Figure~\ref{fig:chiT} shows $\chi(T)$  at $\mu_0H=$ 5~T for the three orientations. To obtain the intrinsic susceptibility, the impurity contribution extracted from the  $M(H)$ analysis was converted to $\chi_\text{imp}(T)$ and subtracted from the measured $\chi(T)$.
The corrected $\chi(T)$ exhibits a broad maximum at
$T_{\mathrm{max}} \approx 250$ K, which is a hallmark of strong short-range
antiferromagnetic correlations in low-dimensional magnetic systems~\cite{Ahmed2017,Arjun2019,Mukharjee2019,Mukharjee2020}. Upon cooling below
$T_{\mathrm{max}}$, the susceptibility decreases gradually as antiferromagnetic
correlations strengthen. At lower temperatures, a clear anomaly appears at
$T_\mathrm{N} \approx 20$ K, marking the onset of long-range antiferromagnetic order. Below
$T_\mathrm{N}$, the susceptibility becomes anisotropic, consistent with an easy-axis ordered
state with the ordered moments aligned along the $b$-axis, in agreement with the
spin-flop transition observed in the magnetization data.

To quantify the dominant intrachain exchange interaction, the intrinsic susceptibility
in the paramagnetic regime was analyzed using the Fisher model for a classical
Heisenberg antiferromagnetic chain~\cite{Fisher1964}. In this model the susceptibility
is given by

\begin{equation}
\chi(T)=\frac{N_A g^2\mu_B^2 S(S+1)}{3k_BT}\frac{1+u}{1-u},
\end{equation}

where

\begin{equation}
u=\coth\!\left(\frac{-2J_F S(S+1)}{k_BT}\right)+\frac{k_BT}{2J_F S(S+1)} .
\end{equation}

Here $J_F$ is defined in the Fisher Hamiltonian
$\mathcal{H}=2J_F\sum_i \mathbf{S}_i\cdot\mathbf{S}_{i+1}$, where
antiferromagnetic coupling corresponds to $J_F>0$. The susceptibility data were fitted in the temperature range 200--400~K, well above $T_\mathrm N$, where 3D critical fluctuations are absent and the response is dominated by intrachain correlations, allowing a 1D chain description. The fit yields a
Land\'e $g$-factor $g = 2.1(3)$ and an effective intrachain exchange interaction
$J \approx $140~K (in the conventional Hamiltonian
$\mathcal{H}=J\sum_i \mathbf{S}_i\cdot\mathbf{S}_{i+1}$). 
We retain the Fisher analysis as a useful classical reference; however, because the magnetic ions carry \(S=3/2\), a quantum treatment provides a more appropriate comparison with the experimental susceptibility.

To assess the influence of the quantum \(S=3/2\) character, we additionally calculated the susceptibility of the nearest-neighbor Heisenberg chain, $\mathcal{H}_{\rm NN}=J\sum_i \mathbf{S}_i\cdot\mathbf{S}_{i+1}$,
by exact diagonalization of periodic chains with \(N=6\) and \(N=7\) sites. The Hamiltonian was diagonalized in sectors of fixed total \(S^z\), and the zero-field susceptibility was obtained from the thermal fluctuations of the total magnetization,
\begin{equation}
\chi(T)=
\frac{N_A(g\mu_B)^2}{Nk_BT}
\left\langle(S_{\rm tot}^{z})^2\right\rangle.
\end{equation}
To reduce finite-size and even-odd effects, the \(N=6\) and \(N=7\) susceptibilities were averaged. A simultaneous fit of the three crystallographic orientations over \(200\leq T\leq380\) K, using a common \(J\) and orientation-dependent \(g\) factors, yields $J/k_B\simeq114~{\rm K},$
with \(g_a\simeq2.037\), \(g_b\simeq2.033\), and \(g_{c^*}\simeq2.037\). The quantum calculation therefore gives a somewhat smaller characteristic exchange scale than the classical Fisher description. Because the calculation employs finite chains, the extracted value should be regarded as a finite-size high-temperature estimate rather than a thermodynamic-limit determination.

Motivated by the AF–FM zig-zag-chain exchange topology proposed in Refs. [30,31], we also considered
\begin{equation}
\mathcal H_{\rm zz}
=
J_1\sum_i{\bf S}_i\cdot{\bf S}_{i+1}
+
J_2\sum_i{\bf S}_i\cdot{\bf S}_{i+2}.
\end{equation}
The susceptibility alone does not constrain \(J_1\) and \(J_2\) independently with sufficient accuracy. We therefore fixed their ratio to \(J_2/J_1=-0.148\), corresponding to the relative AF–FM exchange strengths obtained in Ref. [31], and fitted the overall exchange scale. The resulting constrained fit gives
$
J_1/k_B\simeq102~{\rm K},$ and $J_2/k_B\simeq-15~{\rm K}.
$
The agreement with the susceptibility shows that the experimental data are compatible with the proposed AF–FM zig-zag topology. However, because \(J_2/J_1\) is externally constrained, the susceptibility does not independently establish the microscopic zig-zag exchange parameters.

Although AgCrP\(_2\)S\(_6\) is structurally layered, a conventional isotropic two-dimensional Heisenberg model is not an appropriate representation of the proposed microscopic exchange network. The exchange hierarchy reported in Ref. [31] is strongly anisotropic within the layer, with the dominant interactions following the chain direction and substantially weaker couplings between neighboring chains. We therefore describe AgCrP\(_2\)S\(_6\) as a system of weakly coupled chains embedded within a layered van der Waals structure rather than as an isotropic two-dimensional antiferromagnet.

Taken together, the three analyses consistently identify a dominant antiferromagnetic energy scale of order \(100\) K. All descriptions support predominantly quasi-one-dimensional antiferromagnetic interactions. Deviations from a simple isolated-chain description on cooling may reflect the combined influence of additional intrachain exchange pathways, interchain coupling, and magnetic anisotropy.

The hierarchy of energy scales $J \gg T_N$ indicates that strong short-range
correlations develop well above the magnetic transition, implying that correlated spin
dynamics persist far into the paramagnetic regime. This large separation of scales already suggests that  AgCrP$_2$S$_6$ resides in an extended crossover regime, making it an excellent candidate for a local-probe study of dimensional crossover in a layered vdW magnet. To probe the evolution of these
local spin correlations and fluctuations across this crossover, we turn to $^{31}$P NMR
measurements, which provide a sensitive microscopic probe of both static and dynamic
spin correlations.

\subsection{$^{31}$P Nuclear magnetic resonance (NMR)}
\subsubsection{$^{31}$P NMR spectra and shift}
\begin{figure}
  \centering
  \includegraphics[clip,width=1\columnwidth]{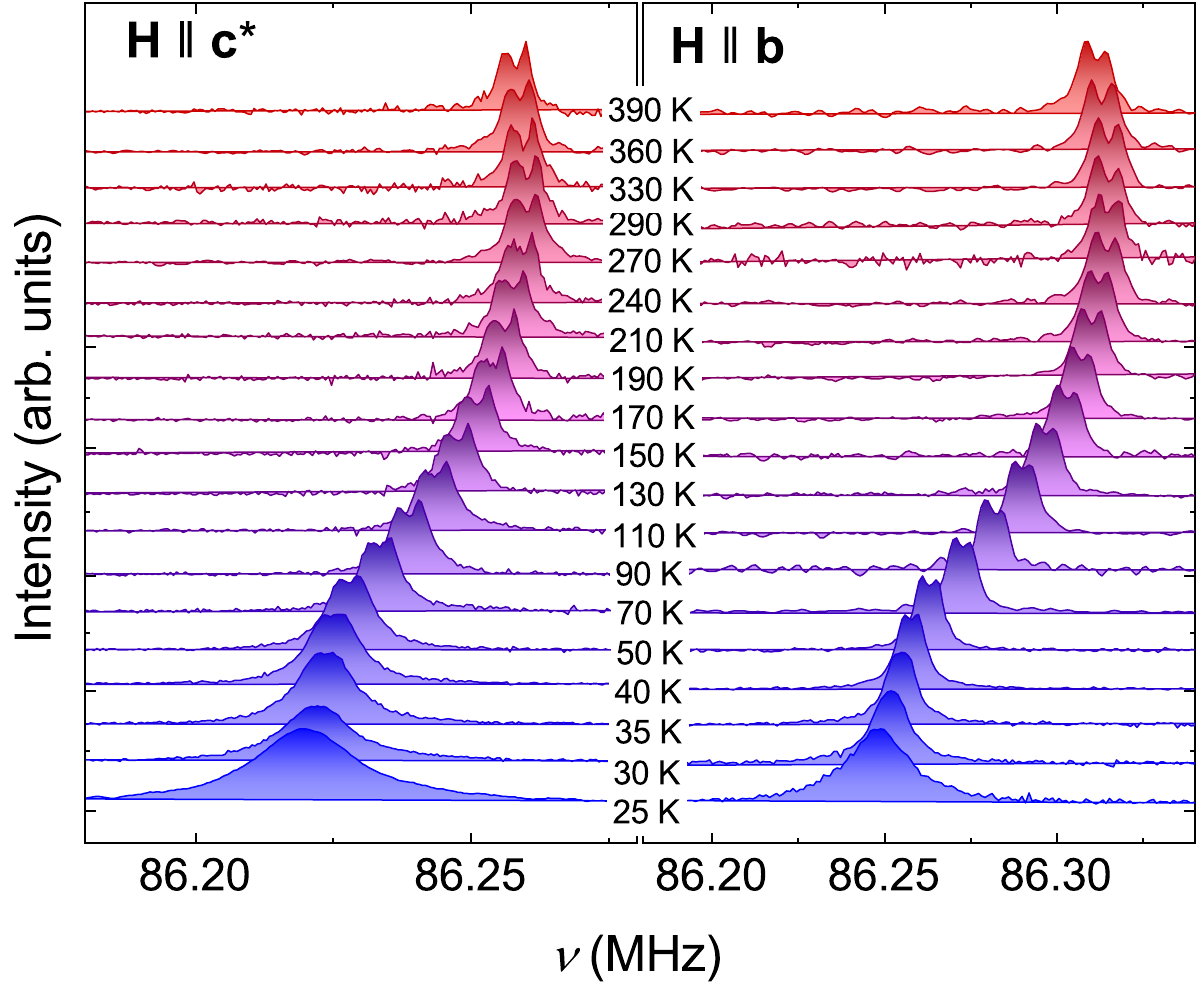}
  \caption{ $^{31}$P NMR spectra of AgCrP$_2$S$_6$ recorded at selected
temperatures for magnetic fields ($\mu_0H=5$ T) applied along the $c^{*}$-axis
 and along the $b$-axis.}\label{fig:NMRspectra}
\end{figure}

Figure~\ref{fig:NMRspectra} presents representative $^{31}\text{P}$ NMR  spectra acquired at various temperatures in an external field of $\mu_0H_0 = 5\text{T}$ applied along the crystallographic $c^*$ and $b$ axes. At higher temperatures in the paramagnetic regime, the $^{31}$P spectra exhibit a resolved double-peaked line shape. The separation of the two components is nearly temperature independent, whereas the overall magnetic linewidth increases progressively on cooling toward $T_\mathrm{N}$. As a consequence, the two components become increasingly difficult to resolve close to the magnetic transition.

The observed double-peaked line shape corresponds to a Pake doublet arising from
the secular dipolar interaction between the two $^{31}$P nuclei forming the
structural P--P dimer within the $\mathrm{P}_2\mathrm{S}_6$ unit~\cite{Pake1948}.
The angular dependence of the dipolar interaction follows the characteristic
$(3\cos^2\theta-1)$ form, where $\theta$ is the angle between the P--P bond
and the applied magnetic field, producing the well-known double-peaked line shape
pattern. Similar dipolar-induced doublets have been reported in related
thiophosphate compounds such as NiPS$_3$~\cite{Dioguardi2020,Bougamha2022}, where the intradimer
P--P geometry leads to a robust temperature-independent splitting.
The temperature-independent Pake splitting observed here confirms the
rigidity of the P--P dimer over a wide temperature range.

\begin{figure}
  \centering
  \includegraphics[clip,width=1\columnwidth]{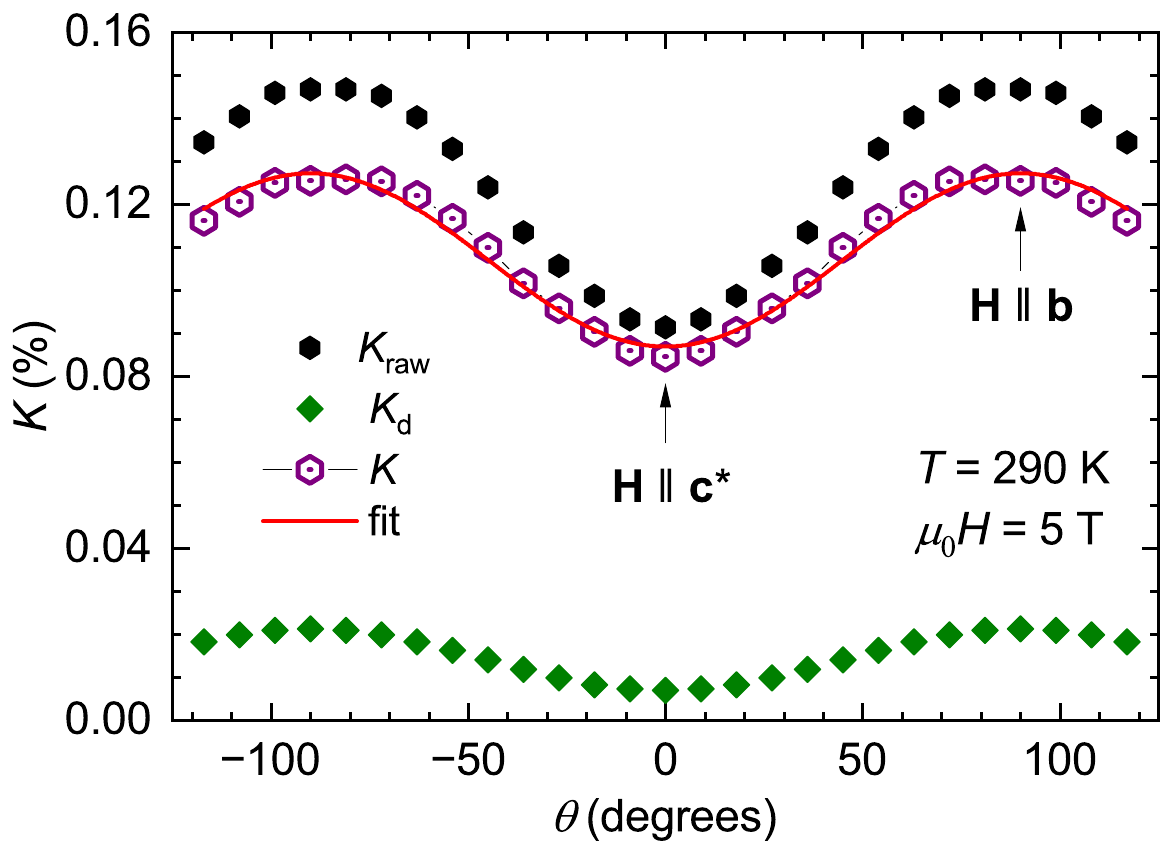}
  \caption{Experimental orientation dependence of  NMR shift between  ($\theta = 0^\circ$) and ($\theta = 90^\circ$). The corrected shift \textit{K} = $K_{\rm raw}$ - $K_{\rm d}$, where $K_{\rm raw}$ is the calculated shift from spectra observed and $K_{\rm d}$ is the contribution from bulk magnetic effects.}\label{fig:rotation}
\end{figure}

\begin{figure*}
  \centering
  \includegraphics[clip,width=2\columnwidth]{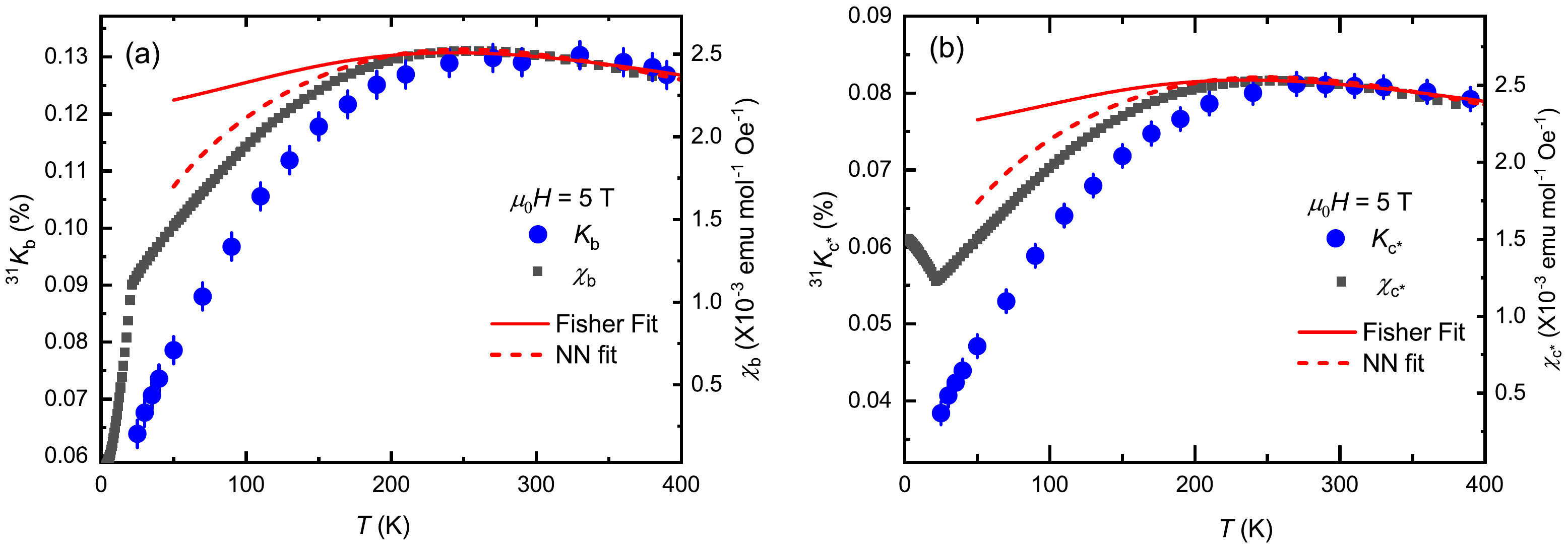}
  \caption{Temperature dependence of the $^{31}$P NMR shift in AgCrP$_2$S$_6$ measured at $\mu_0H_0=5$~T for (a) $H\parallel b$ and (b) $H\parallel c^{*}$. The solid line represents the fit using the Fisher model. }\label{fig:shift}
\end{figure*}

To determine the anisotropy of the shift tensor, the spectra were measured
while rotating the sample out of the $ab$ plane at 300 K. The angular
dependence of the resonance position between $H\parallel c^{*}$
($\theta=0^{\circ}$) and $H\parallel ab$ ($\theta=90^{\circ}$) is shown
in Fig.~\ref{fig:rotation}. The measured raw shifts $K_{\mathrm{raw}}$
were corrected for demagnetization effects arising from bulk magnetization
using $K=K_{\mathrm{raw}}-K_d$, where $K_d$ accounts for macroscopic field
contributions determined by the sample geometry and the angular dependence
of the susceptibility.

For an anisotropic system the angular dependence of the NMR shift can
be written as
\begin{equation}
K(\theta)=K_{\mathrm{iso}}+K_{\mathrm{ax}}
\left[3\cos^2(\theta-\theta_0)-1\right].
\end{equation}
Following the behavior observed in related thiophosphate compounds~\cite{Bougamha2022,Saramgi2026}, we approximate the in-plane components of the $^{31}$P shift as equivalent, $K_a\simeq K_b\equiv K_{ab}$. In particular, in the closely related compound CuCrP$_2$S$_6$, the in-plane variation of the $^{31}$P shift is only about 0.01\%, considerably smaller than the difference between the in-plane and $c^*$ directions. This supports the use of an approximately axial shift tensor for the present analysis. Within this approximation, the isotropic component is $K_{\mathrm{iso}}=(2K_{ab}+K_{c^{*}})/3$. Fitting the angular dependence
yields the principal components of the Knight-shift tensor
$K_{c^{*}}=0.086(4)\%$ and $K_b=0.131(6)\%$, indicating moderate
anisotropy of the hyperfine field at the phosphorus site.

\begin{figure*}
  \centering
  \includegraphics[clip,width=2\columnwidth]{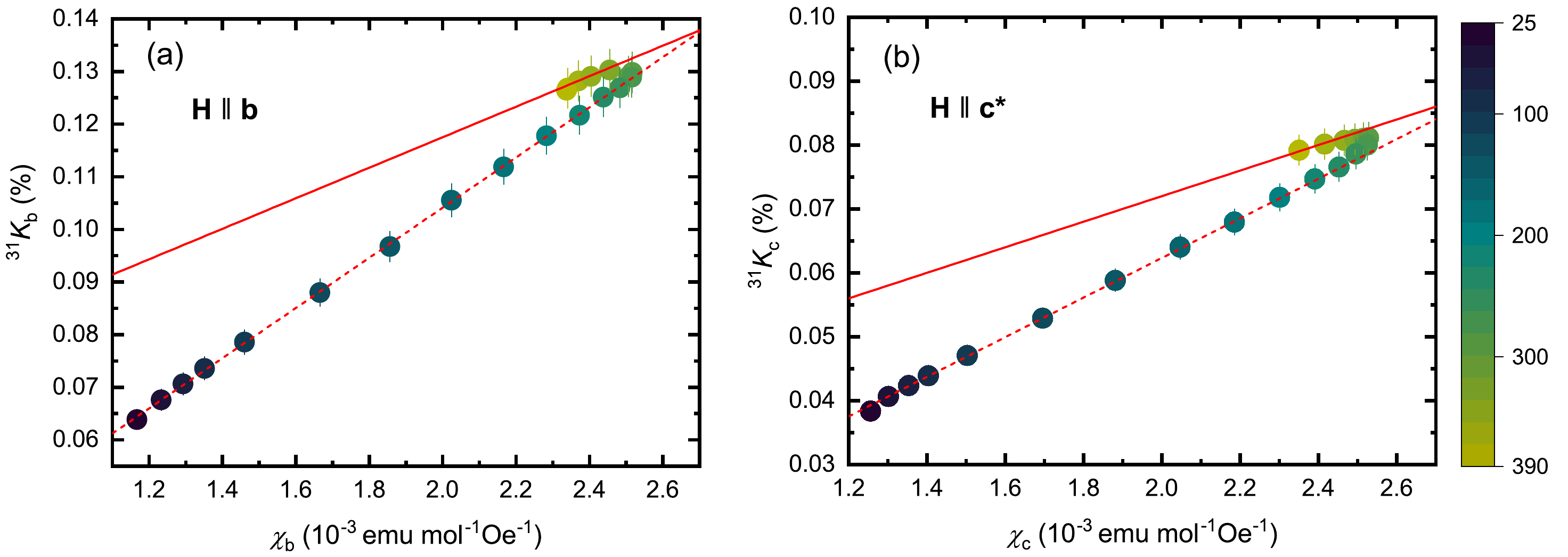}
  \caption{$^{31}$P NMR shift $K$ as a function of bulk susceptibility $\chi$ (Clogston--Jaccarino plots) with temperature as an implicit parameter for (a) $H\parallel b$ and (b) $H\parallel c^{*}$. the solid and dashed lines are linear fits in the high- and low-temperature regimes, respectively, used to extract the hyperfine couplings and intercepts.}\label{fig:KChi}
\end{figure*}

Temperature dependence of NMR shift $K(T)$ is plotted in Fig.~\ref{fig:shift} together with $\chi(T)$ for comparison. At high temperatures, the $K(T)$ follows the bulk magnetic
susceptibility, indicating that the shift originates predominantly from
hyperfine coupling to the Cr$^{3+}$ moments. Notably, $K(T)$ also exhibits a well-defined broad maximum near 
250~K, similar to $\chi(T)$, indicating that both quantities capture the onset of short-range antiferromagnetic correlations at high temperatures. However, below~250 K, a clear deviation develops between the two quantities: while both decrease on cooling, the NMR shift decreases more rapidly than the bulk susceptibility. This difference shows that the local static spin response sensed by the $^{31}$P nucleus evolves differently from the macroscopic uniform susceptibility below 250~K. The NMR shift is related to the intrinsic susceptibility by
\begin{equation}
K(T)=K_0+\frac{A_{\mathrm{hf}}}{N_A\mu_B}\chi(T),
\end{equation}
where $K_0$ is the fit intercept and
$A_{\mathrm{hf}}$ is the hyperfine coupling constant.
The resulting $K$--$\chi$ plot (Fig.~\ref{fig:KChi}) reveals two
distinct linear regimes separated near $T^{*}\approx250$~K,
with clearly different slopes in the high- and low-temperature regions.
This crossover temperature coincides with the broad maxima in the susceptibility. Similar behavior has been also reported for Mn$_2$P$_2$S$_6$ and Ni$_2$P$_2$S$_6$~\cite{Dioguardi2020,Bougamha2022}. 
The effective hyperfine couplings and fit intercepts extracted from linear fits to the low- and high-temperature $K$--$\chi$  regimes are summarized in Table~\ref{Hyp}.

\begin{table}
\caption{Hyperfine coupling constants $A_{\rm hf}$  and fit intercepts $K_0$ extracted from fits to $K$ versus $\chi$ for the field 5~T applied along $b$ and $c^*$ axes:  }
\begin{tabular}{ccc}
  \hline  \hline
& Low temperature&High temperature \\  \hline 
$A_{\rm b}$ (${T/\mu_{\rm B}}$)      &   0.26     &   0.17\\
$A_{\rm c^*}$ (${T/\mu_{\rm B}}$)      &   0.17     &   0.11\\
$K_{\rm 0,b}$ (\%)      &   0.009      &   0.059\\
$K_{\rm 0,c^*}$ (\%)      &   0.0003      &   0.032\\ \hline\hline

  \hline
\end{tabular}
\label{Hyp}
\end{table}

To further test the consistency between the microscopic NMR probe and the bulk magnetic response, the temperature dependence of \(K(T)\) was calculated using both the classical Fisher-chain susceptibility with \(J/k_B\approx140\) K and the quantum nearest-neighbor susceptibility with \(J/k_B\approx114\) K introduced in the susceptibility section (Fig.~\ref{fig:shift}). Using the high-temperature hyperfine coupling constant obtained from the \(K\)–\(\chi\) analysis, with the exchange parameters fixed to the values determined independently from the bulk susceptibility, the calculated NMR shift reproduces the experimental \(K(T)\) data well in the high-temperature regime above \(T^*\). This agreement confirms that \(^{31}\)P NMR reliably tracks the intrinsic spin susceptibility of the Cr\(^{3+}\) chains in this temperature range.

\subsubsection{Ordered-state $^{31}$P NMR spectra and internal field}
\label{subsec:ordered_nmr}
\begin{figure}
  \centering
  \includegraphics[clip,width=1\columnwidth]{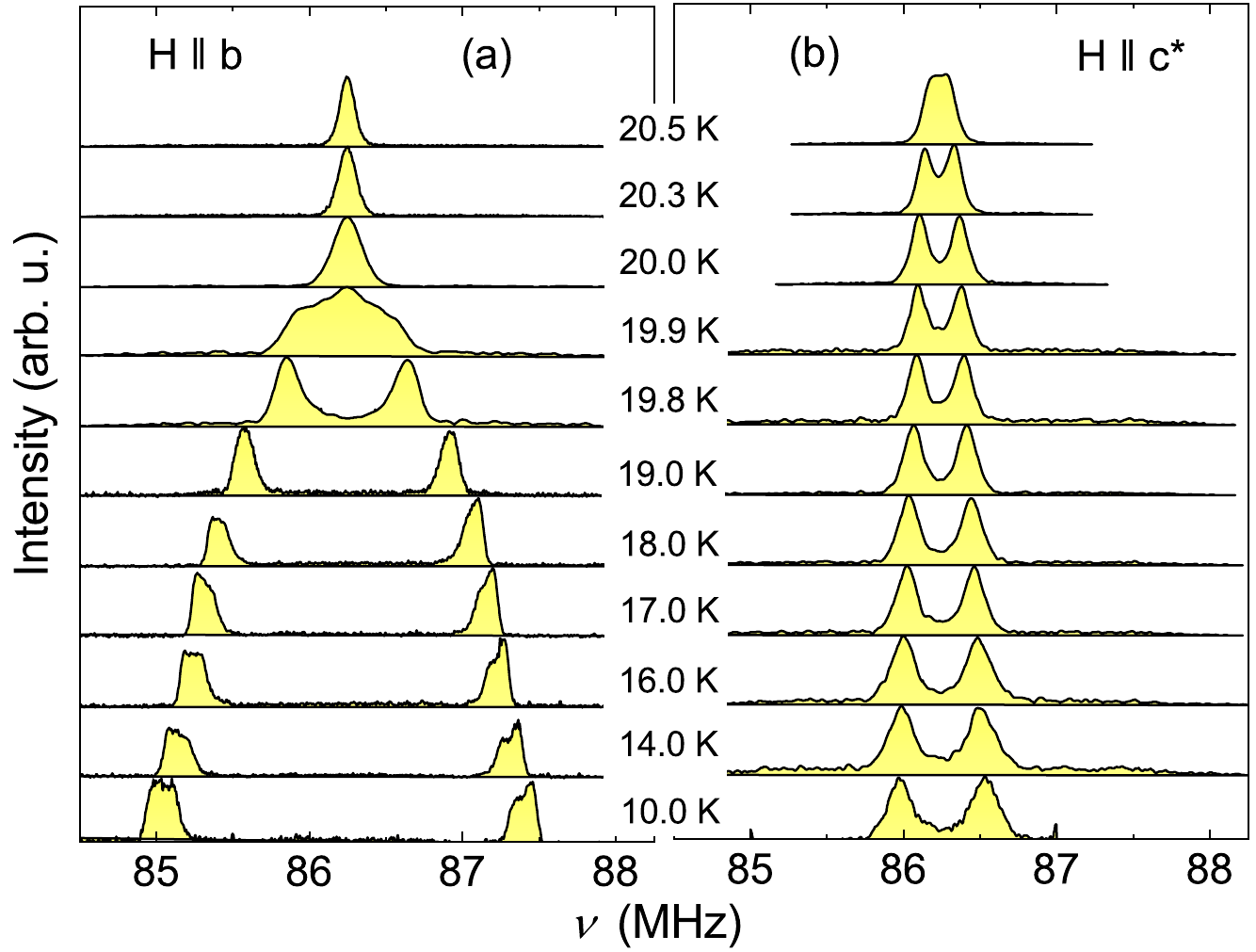}
  \caption{ $^{31}$P NMR spectra of AgCrP$_2$S$_6$ measured at representative temperatures
below $T_\mathrm{N}$ for (a) $H\parallel b$ and (b) $H\parallel c^{*}$, showing the development
of a two-peak structure in the ordered state.}\label{fig:NMR_ordered}
\end{figure}

Figure~\ref{fig:NMR_ordered} shows the evolution of the $^{31}$P NMR spectra on
cooling through the N\'eel temperature $T_\mathrm{N} \approx 20$~K. In the ordered state
($T<T_N$), the paramagnetic resonance develops a clear splitting into two peaks,
indicating the appearance of a static internal hyperfine field at the $^{31}$P site. The splitting onsets at $T \approx 19.9$~K ($H\parallel b$) and $T \approx 20.5$~K ($H\parallel c^{*}$), which are consistent within the experimental temperature uncertainty. For $H\parallel b$, the spectra consist of two well-defined symmetric peaks whose
separation increases continuously with decreasing temperature. For $H\parallel c^{*}$,
the spectra evolve into a broadened two-peak structure with a reduced separation,
reflecting the projection of the internal field onto the external field direction.

The magnitude of the internal field, which is proportional to the Cr$^{3+}$ sublattice magnetization, was obtained from the line splitting $\Delta\nu$ for $H\parallel b$ as
\begin{equation}
H_{\rm int}(T)=\frac{\Delta\nu(T)}{2\gamma_n},
\label{eq:Hint_extract}
\end{equation}
where $\gamma_n$ is the gyromagnetic ratio of $^{31}$P.

\begin{figure}
  \centering
  \includegraphics[clip,width=1\columnwidth]{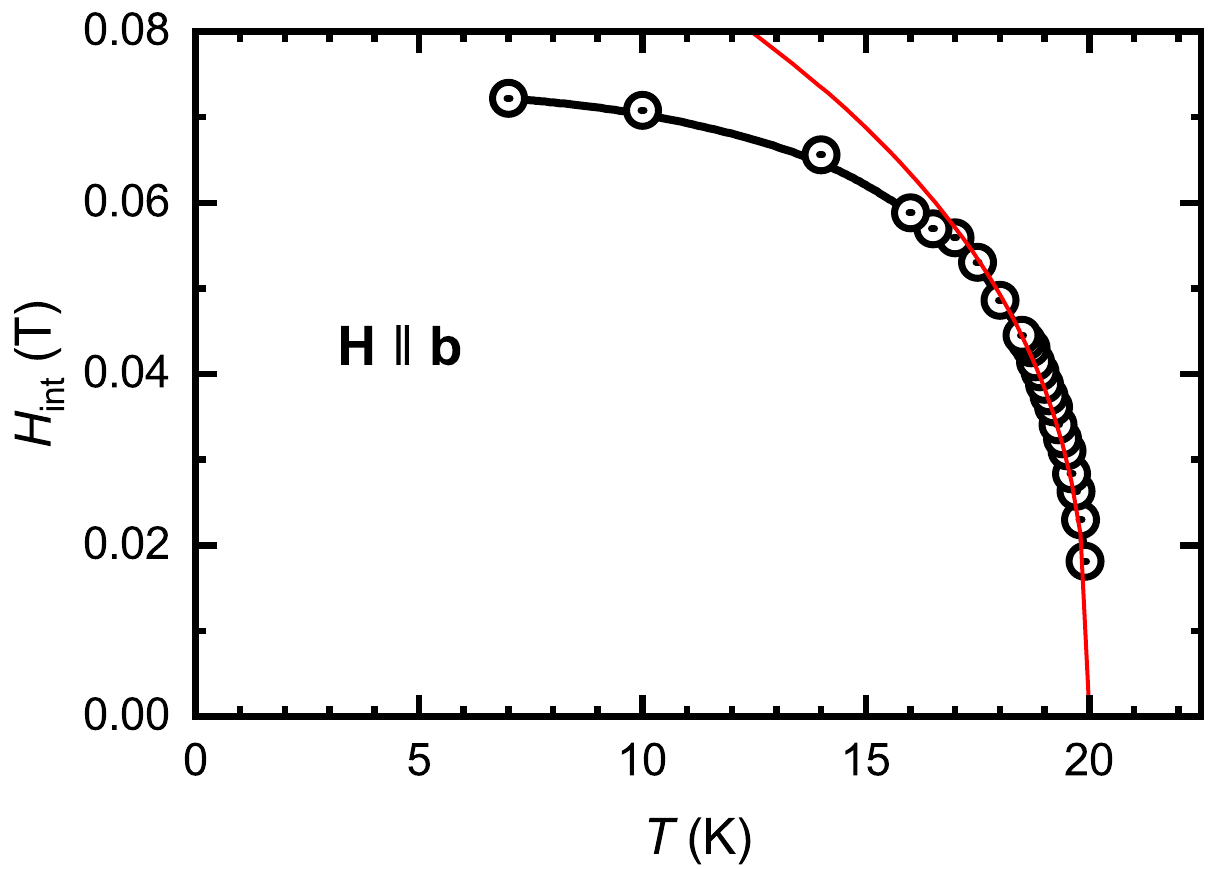}
  \caption{Temperature dependence of the internal hyperfine field $H_{\rm int}(T)$ extracted
from the peak separation using Eq.~\eqref{eq:Hint_extract}. The solid line is a fit to
Eq.~\eqref{eq:Hint_beta}.}\label{fig:Hint}
\end{figure}

The resulting $H_{\rm int}(T)$ is plotted in Fig.~\ref{fig:Hint}. Upon warming, $H_{\rm int}(T)$ decreases rapidly and vanishes at $T_\mathrm{N} $, consistent with an order-parameter-like onset of long-range antiferromagnetic order. Close to the transition, the internal field can be described by the power-law form
\begin{equation}
H_{\rm int}(T)=H_0\left(1-\frac{T}{T_\mathrm{N} }\right)^{\beta},
\label{eq:Hint_beta}
\end{equation}
where $\beta$ is the critical exponent. To estimate $\beta$, we fitted the data in the immediate vicinity of $T_\mathrm{N} $ using temperature steps of 0.1~K (Fig.~\ref{fig:Hint}). A fit over the range 18.5--20.0~K yields $\beta \approx 0.36(2)$. This value is close to those expected for 3D universality classes and is consistent with anisotropic 3D critical behavior~\cite{Pelissetto2002,Nath2009,Ranjith2015}.

 \subsubsection{Spin--lattice relaxation rate }
\label{subsec:T1}

\begin{figure}
  \centering
  \includegraphics[clip,width=1\columnwidth]{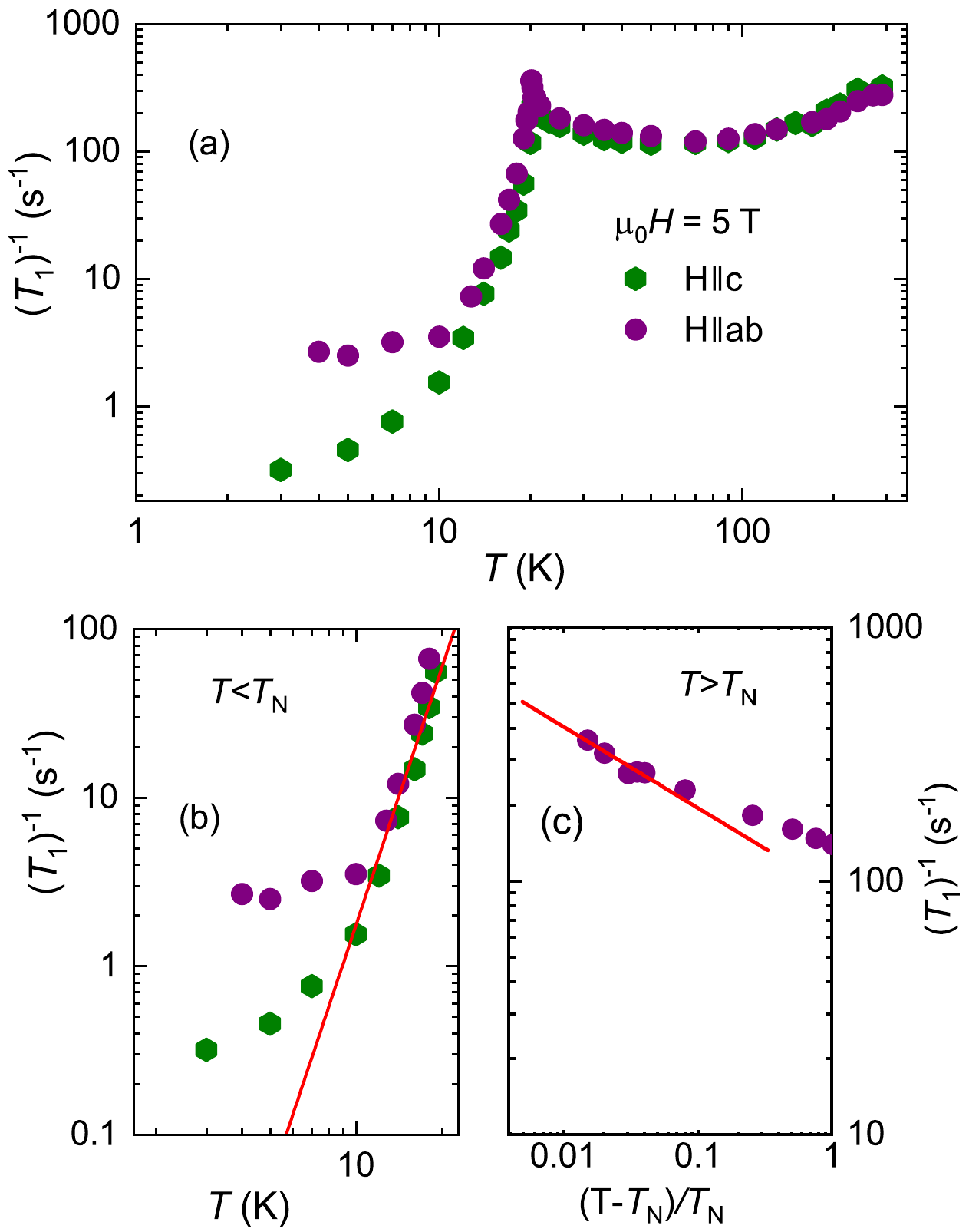}
  \caption{
Temperature dependence of the $^{31}$P spin--lattice relaxation rate $1/T_1$
measured at $\mu_0H=5$ T for $H\parallel c^*$ and $H\parallel b$.
(a) $1/T_1$ over the full temperature range.
(b) Low-temperature region showing the rapid decrease of $1/T_1$
below the N\'eel temperature.
(c) $1/T_1$ above $T_\mathrm{N}$ plotted as a function of
$(T-T_\mathrm{N})/T_\mathrm{N}$.
The solid lines represent fits to the critical and power-law behaviors
described in the text.}\label{fig:T1}
\end{figure}

Since $^{31}$P has the nuclear spin $I = \frac{1}{2}$, the value of $T_1$ at each temperature was estimated by fitting the recovery curve of the longitudinal magnetization to a single exponential function 
\begin{equation}
\label{spinhalf}
  \left(1-\frac{M(t)}{M(\infty)}\right)=A~\mathrm{exp}(-\frac{t}{T_1}).
\end{equation}
Here, $M(t)$ is the nuclear magnetization at a time $t$ after the saturation pulse and $M(\infty)$ is the equilibrium magnetization.
The temperature dependence of the $^{31}$P spin--lattice relaxation rate $1/T_1$
measured at $\mu_0H = 5$~T for magnetic fields applied along the $c^{*}$ and $b$ axes is shown in Fig.~\ref{fig:T1}. Over the full temperature
range the relaxation rate exhibits distinct behaviors in the paramagnetic and
ordered regimes.

At high temperatures ($T \gtrsim 250$~K), $1/T_1$ is nearly temperature
independent for both field orientations, with values of approximately
300~s$^{-1}$. The weak temperature dependence is consistent with exchange-narrowed paramagnetic spin fluctuations.

Upon cooling from room temperature, $1/T_1$ decreases below $T^{*}\sim 250$~K for both field orientations and follows $1/T_1 \propto T^{0.5}$ over $60$--$240$~K and reaches a broad minimum near $\sim 50$~K. On further cooling, $1/T_1$ turns upward and increases rapidly toward $T_\mathrm{N}\approx 20$~K, producing a pronounced peak-like enhancement in the vicinity of the transition. This behavior reflects the slowing down of spin fluctuations as long-range antiferromagnetic order is approached.

Below $T_\mathrm{N}$, the relaxation rate decreases rapidly with decreasing temperature
as the system enters the magnetically ordered state. In the intermediate-temperature range $10~\text{K} \lesssim T < T_N$, $1/T_1$ follows a
$T^5$ power-law decrease for both field orientations, although the
temperature dependence is somewhat stronger for $H \parallel c^{*}$. At the lowest
temperatures the relaxation rate tends to flatten and becomes nearly
temperature independent.

\section{Discussion}

The combined magnetization, susceptibility, and $^{31}$P NMR measurements provide a coherent microscopic picture of the magnetic correlations and spin dynamics in AgCrP$_2$S$_6$. The results reveal a clear hierarchy of energy scales governed by strong intrachain exchange interactions and much weaker interchain coupling, leading to pronounced dimensional crossover behavior.

Low temperature magnetization measurements demonstrate clear magnetic anisotropy in AgCrP$_2$S$_6$. While the magnetization increases nearly linearly with magnetic field for $H \parallel a$ and $H \parallel c^*$, the curve for $H \parallel b$ exhibits a pronounced change in slope near $\mu_0H_{\rm sf} \approx$ 6.5~T. This anomaly corresponds to a spin-flop transition characteristic of a uniaxial antiferromagnet and identifies the $b$-axis as the magnetic easy axis of the ordered state. In a weakly anisotropic antiferromagnet the spin-flop field is given by

\begin{equation}
H_{\rm sf} \approx \sqrt{2H_EH_A},
\end{equation}

where $H_E$ and $H_A$ denote the exchange and anisotropy fields, respectively. The exchange field can be estimated from the susceptibility-derived exchange scale, $J \approx 114$~K using

\begin{equation}
H_E = \frac{2zJS}{g\mu_B},
\end{equation}

where $z$ is the number of nearest neighbors connected by $J$. Taking $z = 2$, appropriate for the dominant intrachain coupling, the exchange field is estimated to be $H_E \sim 5\times10^2$~T. Combining this value with the observed spin-flop field (6.5~T) yields an anisotropy field $H_A \approx 0.042$~T. The corresponding single-ion anisotropy constant obtained from $H_A = 2DS/(g\mu_B)$ is $D \approx 0.019$~K, giving $D/J \sim 1.68\times10^{-4}$ and indicating that AgCrP$_2$S$_6$ lies close to the Heisenberg limit~\cite{Mironov2023},  consistent with a weak easy-axis anisotropy responsible for the observed spin-flop transition. We note that this estimate reflects an effective anisotropy, which may include contributions from anisotropic exchange, dipolar interactions, and local structural distortions.

It is useful to contrast this weak effective anisotropy with that of the related $S=1$ Haldane chain compound AgVP$_2$S$_6$. In AgVP$_2$S$_6$, NMR shift measurements yielded a nearly axial single-ion anisotropy $D\approx$ 4.5~K~\cite{Takigawa1995,Takigawa1996}, while the spin susceptibility and relaxation are governed by a Haldane gap of order 300~K. In contrast, AgCrP$_2$S$_6$  has a much smaller effective anisotropy, $D \approx 0.015$~K, and instead of forming a gapped singlet state, its half-integer spin chains develop short-range antiferromagnetic correlations that ultimately evolve into 3D magnetic order below $T_\mathrm{N} \approx 20$~K.

The significant separation between the intrachain exchange interaction and the ordering temperature offers a quantitative insight into the dimensionality of magnetic interactions. From the susceptibility analysis using the quantum nearest-neighbor exchange scale \(J/k_B\approx114\)~K and \(T_N\approx20\)~K gives
$
J/T_N\approx5.7.
$
The substantial separation between these two scales indicates that long-range order is strongly suppressed relative to the dominant intrachain interaction, consistent with the quasi-one-dimensional magnetic character.
In quasi-1D Heisenberg antiferromagnets, long-range order emerges when weak interchain coupling $J_\perp$ becomes relevant as the intrachain correlation length $\xi(T)$ grows upon cooling. Within the random-phase approximation the ordering condition can be written as
$J_\perp \chi_{1D}(Q,T_N) \sim 1,$
where $\chi_{1D}(Q,T)$ is the susceptibility of an isolated chain at the antiferromagnetic wave vector $Q$~\cite{Schulz1996,Yasuda2005}. Thus even small interchain interactions can stabilize three-dimensional magnetic order once correlated spin segments along the chains become sufficiently long.
The present results therefore illustrate a dimensional crossover from correlated quasi-1D spin dynamics at high temperatures to 3D antiferromagnetic order at low temperatures. The crossover temperature $T^\ast \approx 250$~K identified from the NMR-shift analysis coincides with the susceptibility maximum, consistent with the development of short-range intrachain correlations in a quasi-1D Heisenberg system. The change in the 
$K-\chi$ slope across $T^*$ further indicates a modification of the effective hyperfine coupling at the $^{31}$P site. In the absence of a structural transition, this behavior is best understood as arising from the evolution of the correlated spin environment, which alters the transferred hyperfine field sensed by the nucleus. In this sense, AgCrP$_2$S$_6$ is notable not only because it is quasi-one-dimensional, but because this crossover occurs within a layered van der Waals host where weak inter chain/layer couplings may in principle be modified by thickness, pressure, or interfacial engineering.

 A related breakdown of $K – \chi$ scaling has been reported in the layered thiophosphates Ni$_2$P$_2$S$_6$ and Mn$_2$P$_2$S$_6$~\cite{Dioguardi2020,Bougamha2022}. In Ni$_2$P$_2$S$_6$, the extracted $K(\chi)$ slopes differ between the high- and low-temperature regimes below $\sim275$ K, while in Mn$_2$P$_2$S$_6$ two distinct linear regimes are observed and the apparent hyperfine coupling even changes sign between them. AgCrP$_2$S$_6$ likewise exhibits two-slope behavior, placing it within the broader class of thiophosphate systems in which the apparent hyperfine coupling evolves across the correlation crossover. In contrast, the linear $K – \chi$ relation remains intact in CuCrP$_2$S$_6$, where ferromagnetic correlations dominate the paramagnetic state, indicating that the effective hyperfine coupling remains essentially temperature independent over the measured temperature range~\cite{Saramgi2026,abraham2025}. 
The contrast between CuCrP$_2$S$_6$ and the antiferromagnetic thiophosphates suggests that the breakdown of $K – \chi$ scaling is associated with the development of finite-$q$ antiferromagnetic correlations rather than being an intrinsic feature of the thiophosphate family. Although both the NMR shift and the bulk susceptibility probe the uniform ($q=0$) spin response, finite-$q$ correlations determine how neighboring magnetic moments are correlated and therefore how their individual hyperfine-field contributions combine at the ligand nucleus. Consequently, the apparent hyperfine coupling extracted from the $K – \chi$ relation evolves as short-range antiferromagnetic correlations develop. A possible microscopic origin is a correlation-induced modification of the vector sum of the hyperfine fields from neighboring magnetic ions, whose magnitude and even sign depend on the geometry and symmetry of the hyperfine coupling network and on the correlated spin arrangement.

NMR-shift analysis at high temperatures yields anisotropic hyperfine coupling constants $A_{b} \approx 0.17$ T/$\mu_\mathrm B$ and $A_{c^*} \approx 0.11$ T/$\mu_\mathrm B$. The dipolar contribution to the hyperfine tensor was calculated using a lattice-sum method, yielding

\begin{equation}
A_{\mathrm{dip}} =
\begin{pmatrix}
0.016 & 0.000 & 0.001 \\
0.000 & 0.019 & 0.028 \\
0.001 & 0.028 & -0.035
\end{pmatrix}
\ \mathrm{T}/\mu_\mathrm B .
\end{equation}


Although the dipolar contribution is much smaller than the total hyperfine coupling, the tensor contains finite off-diagonal elements, indicating that the local hyperfine field at the P site is intrinsically anisotropic.  Subtracting this dipolar part from the total $A_{b}$ and $A_{c^*}$ shows that the transferred (contact/superexchange) contribution along $b$ and $c^*$ is nearly identical ($\sim$ 0.15~T/$\mu_\mathrm B$) within uncertainty, i.e. essentially isotropic, closely analogous to the situation in CuCrP$_2$S$_6$ where the transferred coupling is isotropic and the NMR shift anisotropy originates primarily from the dipolar field~\cite{Saramgi2026}.

The spin--lattice relaxation rate provides insight into the spin dynamics. At high temperatures the relaxation rate is nearly temperature independent, indicating exchange-narrowed spin fluctuations. In this regime the relaxation rate can be described by Moriya's theory \cite{Moriya1956,Moriya1963},

\begin{equation}
\left(\frac{1}{T_1}\right)_{T\rightarrow\infty} =
\frac{(\gamma_{\mathrm{N}} g\mu_{\rm B})^{2}\sqrt{2\pi}z^\prime S(S+1)}{3\,\omega_{\mathrm{ex}}}
{\Big(\frac{A_{\perp}}{z'}\Big)^{2}},
\label{t1inf}
\end{equation}
where $S$  is the local moment spin, $\gamma_{\mathrm{N}}$ is the nuclear gyromagnetic ratio, $g$ is the electron $g$-factor, $A_{\perp}$ is the hyperfine coupling perpendicular to the applied field, $z^\prime$ is the number of nearest-neighbor Cr$^{3+}$ spins for a given P site, and $\omega_{\mathrm{ex}}$ is the Heisenberg exchange frequency, describes how fast the local fields fluctuate, set by the exchange motion among spins. The exchange frequency is defined as: 

\begin{equation}
\omega_{ex} =
\frac{Jk_B}{\hbar}
\sqrt{\frac{2zS(S+1)}{3}},
\end{equation}
where $z$ is the number of nearest-neighbor spins of each Cr$^{3+}$ ion and $J_{\rm max}$ is the nearest-neighbor Heisenberg exchange constant.
In AgCrP$_2$S$_6$, the two $^{31}$P nuclei forming the structural dimer couple to the same set of Cr$^{3+}$ spins and their local hyperfine fields fluctuate in a correlated manner. Therefore the single-site  Moriya estimate will be modified. In this situation the nuclear relaxation is described by symmetric and antisymmetric relaxation modes with rates $\lambda_\pm = R_0 \pm R_c$, where $R_c$ represents cross-correlations of the local fields~\cite{Beeman1968}. For strongly correlated fluctuations ($R_c\sim R_0$), the relaxation rate approaches $1/T_1\approx2R_0$.

Using the quantum nearest-neighbor exchange scale \(J/k_B\approx114\) K, \(g_b\simeq2.03\), and \(A_{\rm hf}\approx0.17\) T/\(\mu_B\) yields a single-site estimate \(R_0\sim40\ {\rm s}^{-1}\). Even allowing for correlated fluctuations of the two \(^{31}\)P nuclei forming the structural dimer increases this estimate only to \(\sim80\ {\rm s}^{-1}\), still substantially smaller than the experimentally observed value of \(\sim300\ {\rm s}^{-1}\).
This large discrepancy indicates that the relaxation cannot be described solely by the scalar $q = 0$ hyperfine coupling inferred from the Knight shift. 
Instead, the dominant contribution to $1/T_1$ arises from the full anisotropic hyperfine form factor, including transverse and possibly off-diagonal tensor elements, together with finite-$q$ spin fluctuations. 
In this sense, the Knight shift probes the static uniform spin response, whereas $1/T_1$ samples low-frequency spin fluctuations integrated over momentum space and weighted by the anisotropic hyperfine form factor.

As discussed earlier, $1/T_1$ decreases gradually below $T^* \simeq 250$~K, in contrast to the enhancement often expected from the growth of antiferromagnetic correlations. A similar temperature dependence has been reported in several antiferromagnets~\cite{Nath2009,Carretta2002,Tsirlin2011,Ranjith2015}. For Heisenberg antiferromagnetic spin chains, particularly in the spin-$1/2$ case, theory predicts that the contribution of the staggered component ($q=\pm \pi/a$) to the spin-lattice relaxation rate is nearly temperature independent, $1/T_1 \sim T^{0}$, whereas the uniform component ($q=0$) scales linearly with temperature, $1/T_1 \sim T$~\cite{Sandvik1995,Sachdev1994}. The staggered and uniform components are expected to dominate in the low-temperature ($T \ll J$) and high-temperature ($T \sim J$) limits, respectively~\cite{Nath2005}. Although these predictions are not expected to apply quantitatively to AgCrP$_2$S$_6$, they provide a useful qualitative framework for interpreting the present results. The experimentally observed sublinear power-law behavior, $1/T_1 \propto T^{0.5}$ over the range $60~\mathrm{K} \leq T \leq 210~\mathrm{K}$, indicates that AgCrP$_2$S$_6$ is not in a simple asymptotic regime, but rather in a crossover region where both uniform and antiferromagnetic spin fluctuations contribute to the relaxation. 
This sublinear behavior is inconsistent with a purely uniform high-temperature fluctuation regime and instead suggests that finite-$q$ antiferromagnetic fluctuations increasingly contribute to the low-energy spin dynamics upon cooling. 

Taken together, the static ($K$) and dynamic ($1/T_1$) NMR results indicate that below $T^*$, the system enters a crossover regime in which short-range intrachain antiferromagnetic correlations grow, spectral weight shifts toward finite $q$, and the effective hyperfine coupling sensed at the $^{31}$P site evolves. 
This local-probe view of momentum-space redistribution is central to the broader relevance of AgCrP$_2$S$_6$ as a vdW platform for studying dimensional crossover in correlated magnets. 
The intermediate-temperature evolution can also be compared with the recently reported ESR and Raman results~\cite{Lee2025}. Those measurements revealed anomalies in the ESR parameters and in a Raman Fano mode near \(120\) K, together with a redistribution of magnetic spectral weight and appreciable magnetic Raman scattering in the interchain polarization. These observations were interpreted as evidence for a change in the character of the magnetic excitations as next-nearest-neighbor and interchain interactions become increasingly relevant below this temperature. The \(120\)-K scale lies within the broad intermediate regime over which the present \(^{31}\)P relaxation rate exhibits sublinear temperature dependence. Although the NMR data do not establish a direct correspondence between these two observations, the ESR/Raman results provide complementary evidence that the magnetic response evolves progressively between the predominantly intrachain correlated regime and the three-dimensional ordered state.

On further cooling, $1/T_1$ turns upward and increases rapidly toward $T_\mathrm{N}$, showing pronounced critical enhancement in the vicinity of the transition due to the slowing down of spin fluctuations. The relatively high ordering temperature, $T_\mathrm{N} \approx$ 20 K, and the intrachain exchange scale, $J \approx 100$~K, leaves only a narrow temperature window between critical fluctuations near $T_\mathrm{N}$ and finite-temperature corrections at higher temperatures, hindering the clear identification of a distinct Tomonaga-Luttinger-liquid regime without excluding its existence.
The singular behavior around the phase-transition point is characterized by the
power-law dependence of the physical quantities. In particular, $T_1^{-1}$ characterizes the dynamical property of the phase transition. Figure~\ref{fig:T1}(c) shows log-log plot of the nuclear spin lattice relaxation rate $T_1^{-1}$ against the reduced temperature $\epsilon=(T-T_\mathrm{N})/T_\mathrm{N}$. $T_1^{-1}$ is well described by the power law, $T_1^{-1}\propto\epsilon^{-\gamma}$, for 0.01$<\epsilon<$0.1, where $\gamma$ is the critical exponent. We obtained $\gamma=$ 0.32(3) for this fitting range. A mean field theory for a 3D isotropic Heisenberg antiferromagnet leads to $\gamma$ = 1/2~\cite{Moriya1962}. A dynamic scaling theory indicates $\gamma$ = 1/3 for a 3D isotropic Heisenberg model~\cite{Halperin1967,Hohenberg1977}.  For comparison, larger critical exponents have been reported for low-dimensional systems, e.g., $\gamma \sim 0.8$ for quasi-2D Heisenberg systems~\cite{Takeya2008,deJongh2012} and $\gamma \sim 1.5$ for 2D Ising systems~\cite{Oyamada2008,collins1989}. The exponent of $\gamma$ = 0.32(3) suggests that the critical dynamics of AgCrP$_2$S$_6$ are governed by a 3D universality class. Moreover, the critical divergence observed here is confined to a relatively narrow reduced-temperature window, $0.01 < \epsilon < 0.1$, consistent with 3D criticality, whereas much broader critical regimes ($\epsilon \lesssim 7$) are typically expected for two-dimensional systems~\cite{itoh2009,Birgeneau1990}.

In the magnetically ordered state ($T<T_\mathrm{N}$), the strong temperature dependence of 1/$T_1$ is a clear signature of the relaxation due to scattering of magnons by the nuclear spins~\cite{Belesi2006}. For $T\gg\Delta/k_{\rm B}$, 1/$T_1$ follows either a $T^3$ behavior or a $T^5$ behavior due to a two-magnon Raman process or a three-magnon process, respectively, where $\Delta$ is the energy gap in the acoustic magnon spectrum~\cite{Beeman1968}. Corresponding $T^3$ or $T^5$ temperature dependences have been reported for several low dimensional spin systems in the antiferromagnetic ordered phase~\cite{Saramgi2026,Ranjith2022,Ranjith2024}.  In AgCrP$_2$S$_6$, the relaxation rate  follows $1/T_1\propto T^5$ in the ordered state, consistent with magnon-mediated relaxation in antiferromagnets via three-magnon Raman scattering processes.

Finally, the internal hyperfine field extracted from the NMR line splitting follows an order-parameter-like temperature dependence $H_{\rm int}(T)\propto(1-T/T_N)^\beta$ with $\beta\approx0.36(2)$. This value is close to the critical exponents expected for 3D universality classes, including the 3D Heisenberg ($\beta\approx0.365$) and 3D XY ($\beta\approx0.345$) models~\cite{Pelissetto2002,Nath2009,Ranjith2015}. The critical behavior therefore confirms that the ordered state is governed by 3D interactions despite the strong quasi-1D correlations present at higher temperatures.

The observation of a line splitting below $T_{\rm N}$ demonstrates the presence of a finite static internal hyperfine field at the P site in the antiferromagnetically ordered state. This is fully compatible with the proposed interpretation of the $K – \chi$ anomaly above $T_{\rm N}$. The evolution of the apparent hyperfine coupling in the paramagnetic regime reflects how developing short-range antiferromagnetic correlations modify the superposition of the hyperfine-field contributions from neighboring magnetic ions. Upon magnetic ordering, these correlations become static, giving rise to the finite internal hyperfine field responsible for the observed spectral splitting. The ordered-state splitting therefore does not represent a separate phenomenon but rather the continuation of the correlation-driven evolution of the hyperfine interaction across the magnetic transition.

\section{Conclusion}

We report a comprehensive investigation of the magnetic properties and spin dynamics of the chain-like van der Waals antiferromagnet AgCrP$_2$S$_6$ using magnetization, magnetic susceptibility, and $^{31}$P nuclear magnetic resonance (NMR). The magnetic susceptibility exhibits a broad maximum near $T_{\rm max}\approx250$~K, characteristic of strong short-range antiferromagnetic correlations in low-dimensional systems. The susceptibility is governed by a dominant quasi-one-dimensional antiferromagnetic exchange scale of order \(100\) K. A classical Fisher-chain analysis yields \(J/k_B\approx140\) K, while a quantum \(S=3/2\) nearest-neighbor-chain calculation gives \(J/k_B\approx114\) K. A constrained AF–FM zig-zag-chain calculation is likewise compatible with the experimental susceptibility, yielding \(J_1/k_B\approx102\) K and \(J_2/k_B\approx-15\) K. Magnetization measurements reveal pronounced magnetic anisotropy with a spin-flop transition near $\mu_0H_{\rm sf}\approx$ 6.5~T for fields applied along the $b$-axis, identifying the easy axis of the ordered state. 

$^{31}$P NMR measurements provide microscopic insight into both static and dynamic spin correlations. The paramagnetic spectra exhibit a Pake doublet arising from dipolar coupling within the structural P–P dimer. NMR-shift analysis yields nearly isotropic transferred hyperfine coupling and a crossover in the effective hyperfine coupling across $T^*$. The combined static and dynamic NMR results provide consistent evidence for a crossover in AgCrP$_2$S$_6$ below $T^*$. Two distinct linear regimes separated near $T^*$ indicate a change in the effective transferred hyperfine coupling. In parallel, the sub-linear power-law behavior $1/T_1 \propto T^{0.5}$ over $60$--$240$~K shows that the system lies in a crossover regime where finite-$q$ antiferromagnetic fluctuations increasingly dominate the low-energy spin dynamics.

Together, these results establish that below $T^*$, the system enters a crossover regime in which short-range intrachain antiferromagnetic correlations grow, spectral weight shifts toward finite $q$, and the effective hyperfine coupling sensed at the $^{31}$P site evolves.

The high-temperature spin-lattice relaxation rate significantly exceeds the exchange-narrowed limit derived from the scalar hyperfine coupling, indicating that relaxation is governed by the full anisotropic hyperfine tensor, including transverse and possibly off-diagonal components, as well as finite-$q$ spin fluctuations. Upon cooling toward the N\'eel temperature $T_\mathrm{N}\approx20$~K the relaxation rate increases due to the growth of spin correlations and exhibits critical enhancement near the transition. Below $T_\mathrm{N}$, the relaxation follows $1/T_1\propto T^5$, consistent with magnon-mediated relaxation in the ordered state.

The temperature dependence of the internal hyperfine field yields a critical exponent $\beta\approx0.36(2)$, consistent with 3D universality. Together these results establish AgCrP$_2$S$_6$ as a quasi-one-dimensional $S=3/2$ Heisenberg antiferromagnet embedded in a layered van der Waals structure, exhibiting a dimensional crossover from correlated intrachain spin dynamics at high temperatures to anisotropic 3D magnetic order at low temperatures.

 The quasi-one-dimensional magnetic character combined with the layered van der Waals structure of AgCrP$_2$S$_6$ suggests several promising  
directions for future investigation. Thickness-dependent studies down 
to few-layer limits could probe the evolution from bulk 3D 
order toward purely one-dimensional behavior, potentially revealing quantum 
critical phenomena as the dimensionality is reduced. The weak interlayer 
coupling may enable construction of heterostructures combining AgCrP$_2$S$_6$ 
with other vdW magnets or two-dimensional materials, offering opportunities 
to engineer magnetic properties through proximity effects and interfacial 
exchange coupling. The weak but well-defined magnetic anisotropy identified here, 
combined with the well-defined spin-flop transition, makes this system 
attractive for exploring field-tunable magnon transport. Furthermore, 
the large ratio $J/T_N \approx 5.7$ and the clear spectral weight transfer 
observed in our NMR measurements establish AgCrP$_2$S$_6$ as a model 
system for testing theories of dimensional crossover in correlated quantum 
magnets. External perturbations such as pressure or chemical substitution 
could tune the interchain coupling strength, enabling systematic studies 
of the crossover from one-dimensional to 3D magnetic behavior.

\section*{Acknowledgment}
K.M.R. acknowledges the financial support by the Deutsche Forschungsgemeinschaft
(DFG, German Research Foundation) under Germany’s Excellence Strategy through the Würzburg-Dresden
Cluster of Excellence on Complexity and Topology in Quantum Matter-ct.qmat (EXC 2147, project-id
390858490). S.A acknowledges "MagTop" project (FENG.02.01-IP.05-0028/23) carried out within the "International Research Agendas" programme of the Foundation for Polish Science co-financed by the European Union under the European Funds for Smart Economy 2021-2027 (FENG). H.-J.G. acknowledges the financial support by the DFG through the Grant No: 534290830.
Y.S. acknowledges financial support by the German Federal Ministry of Education and Research (BMBF) through project GU-QuMat (01DK240008). B.B. acknowledges financial support by the Würzburg-Dresden Cluster of Excellence ct.qmat (EXC 2147, Project-id 390858490), the BMBF through project GU-QuMat (01DK240008), and the Deutsche Forschungsgemeinschaft (DFG, German Research Foundation) through Project No. 447482487 and SFB 1143 (Project-id 247310070).

\end{document}